\documentclass{PoS}

\usepackage{graphicx}
\usepackage{amsfonts}
\usepackage{amsmath}
\usepackage{amssymb}
\usepackage{amsthm}
\usepackage{latexsym}
\usepackage{slashed}
\usepackage{array}
\usepackage{cite}

\title{Higgs boson resonance parameters and the finite temperature phase transition in a chirally invariant Higgs-Yukawa model}

\ShortTitle{A chirally invariant Higgs-Yukawa model}

\author{\speaker{John Bulava}\\
        NIC, DESY Zeuthen\\
        E-mail: \email{john.bulava@desy.de}}

\author{Philip Gerhold\\
				NIC, DESY Zeuthen and Humbolt-Universit\"at zu Berlin\\
				E-mail: \email{philip.gerhold@physik.hu-berlin.de}}

\author{Karl Jansen\\
        NIC, DESY-Zeuthen\\
        E-mail: \email{karl.jansen@desy.de}}

\author{Jim Kallarackal\\
				NIC, DESY-Zeuthen and Humbolt-Universit\"at zu Berlin\\
				E-mail: \email{Jim.kallarackal@physik.hu-berlin.de}}

\author{Attila Nagy\\
        NIC, DESY-Zeuthen and Humbolt-Universit\"at zu Berlin\\
        E-mail: \email{attila.nagy@desy.de}}

\abstract{
\vspace*{-10.5cm}
\begin{flushright}
\texttt{\footnotesize DESY 11-208}\\
\end{flushright}
\vspace*{9.5cm}
We study a chirally invariant Higgs-Yukawa model regulated on a space-time lattice. We calculate Higgs boson resonance parameters and mass bounds for various values of the mass 
of the degenerate fermion doublet. Also, first results on the phase 
transition temperature are presented. In general, this model may be relevant 
for BSM scenarios with a heavy fourth generation of quarks.}

\FullConference{ The XXIX International Symposium on Lattice Field Theory - Lattice 2011\\
July 10-16, 2011\\
Squaw Valley, Lake Tahoe, California}

\begin{document}

The development of a lattice regularization of the Dirac fermion 
bilinear which respects a chiral symmetry at finite lattice 
spacing~\cite{Neuberger:1997fp} suggests that lattice studies of the 
Higgs-Yukawa sector be revisited~\cite{Fodor:2007fn,Gerhold:2007yb}. To this end, we examine a model which 
contains 
a single complex scalar doublet and two fermion fields, the left-handed 
components of which are associated into a doublet. The continuum Lagrangian 
density of our model is therefore
\begin{align}
\mathcal{L} = \frac{1}{2}(\partial_{\mu}\phi)^2 + \frac{1}{2}m^2\phi^2 + 
\lambda\phi^4 + \bar{t} \slashed \partial t + \bar{b} \slashed \partial b + y\bigg( {\bar{t} \atop \bar{b}}\bigg)^{T}_L \phi b_R  + y\bigg( {\bar{t} \atop \bar{b}}\bigg)^{T}_L i\sigma_2\phi^{*} t_R  + h.c. ,
\end{align}
where $t$ and $b$ denote the fermion fields and $\sigma_2$ is the second Pauli matrix. Note that we have set the Yukawa couplings of both fermion fields to be equal. Additionally, we shall find it convenient to define the parameters $\kappa$ and $\hat \lambda$ via 
\begin{align}
\lambda = \frac{ \hat \lambda}{4 \kappa^2}, \qquad m^2 = \frac{1 - 2\hat{\lambda} - 8\kappa }{\kappa}\:.
\end{align}
At finite lattice spacing, the overlap discretization is used, which 
provides well-defined chiral projectors to isolate the left and right 
handed components of the fermion fields. For details on the discretization, 
see Ref.~\cite{Gerhold:2010wy}. Furthermore, it is assumed that we are in the 
broken 
phase where the scalar field acquires a non-zero vacuum expectation value $v$. 
Although there is no explicit fermion mass term in the Lagrangian, in the 
broken phase the fermions acquire a mass ($m_f$) which at leading order in the quartic and 
Yukawa couplings is proportional to $y\times v$.  
Large Yukawa couplings are therefore required to study the model at large $m_f$, where the perturbative expansion may break down. 

Due to the lack of gauge fields (the effects of which presumably can be described perturbatively), the complicated Neuberger-Dirac operator is diagonal in 
momentum space and may therefore be constructed exactly, up to machine 
precision. The model can then be evaluated efficiently using the Fast 
Fourier Transform (FFT) algorithm. 

To simulate the model, we employ the pHMC 
algorithm~\cite{Frezzotti:1997ym}, with many improvements~\cite{Gerhold:2010wy}.
Specifically, Fourier acceleration~\cite{Davies:1985ad,Catterall:2001jg} is 
necessary to reduce the autocorrelation times of low-momentum modes of the 
Higgs and fermion propagators, and several types of preconditioning have been 
used to reduce the condition number of the Dirac operator. 

We set the scale in our theory via the renormalized vacuum expectation value $v_R$ of the scalar field, which is 
non-zero in the broken phase. In the standard model, this can be expressed as    
\begin{align}
v_R = \frac{v}{\sqrt{Z_G}} = \frac{2M_W}{g} = 246\mathrm{GeV}, 
\end{align} 
where $g$ is the weak gauge coupling and 
$Z_G$ is the renormalization constant for the Goldstone boson propagator determined by the condition
\begin{align}
Z_G^{-1}  =  \frac{d}{dp^2} \mathrm{Re}\; ( G_G^{-1}(p^2))\big|_{p^2 = m_{G}^2}.
\end{align}
In practice $Z_G$ is determined from fits to the Goldstone propagator. 

The lattice cutoff in our theory cannot be removed while maintaining non-zero 
interactions. This is the well known triviality (i.e. existence of a Landau pole in perturbation theory) of the $\phi^4$ theory. There 
is some evidence from the large-$N_F$ expansion~\cite{Fodor:2007fn} that the 
theory with fermions is also trivial. 

This model has been used to study the triviality and vacuum instability 
bounds of the Higgs boson at $m_f = m_t = 175\mathrm{GeV}$~\cite{Gerhold:2010bh,Gerhold:2009ub} as well as the phase structure of the theory~\cite{Gerhold:2007gx,Gerhold:2007yb}. 
It has been shown that in the $\phi^4$ theory the Higgs boson mass is an 
increasing function of the bare quartic coupling $\lambda$ at a fixed value 
of the cutoff\cite{Luscher:1987ay}. Furthermore, the quartic coupling must be $\ge 0$ 
to preserve the stability of the theory. It was found for the model considered here~\cite{Gerhold:2010bh} that the maximum value of 
the Higgs boson mass occurs at $\lambda=\infty$ while the minimum occurs 
at $\lambda=0$. 

The Higgs mass is determined by the pole position of the Higgs boson propagator
\begin{align}
\mathrm{Re} \; ( G_H^{-1} (p^2) ) \Big |_{p^2 = -M_H^2} = 0.
\end{align}
We determine the pole position by fitting the propagator to an ansatz which
neglects the finite width of the Higgs to decay into Goldstone bosons. 
The Higgs mass can also be determined from the temporal correlation function 
of two suitable interpolating fields. 

\begin{figure}[ht]
						\centering
						\includegraphics[width=0.31\textwidth]{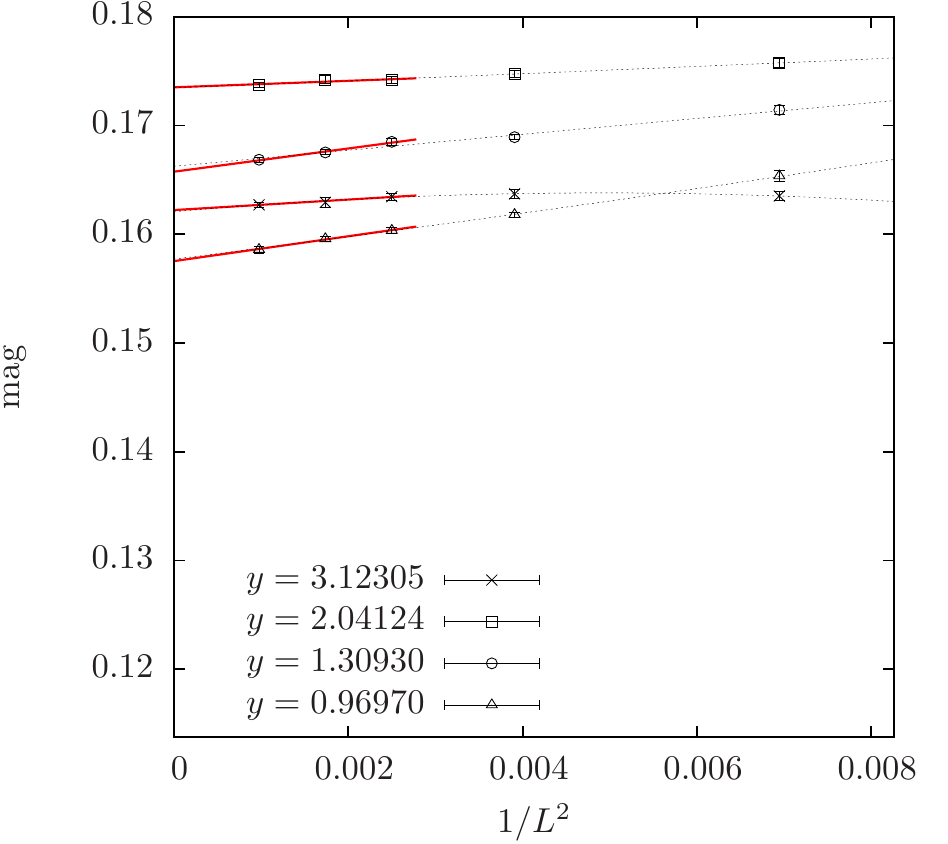}
						\includegraphics[width=0.31\textwidth]{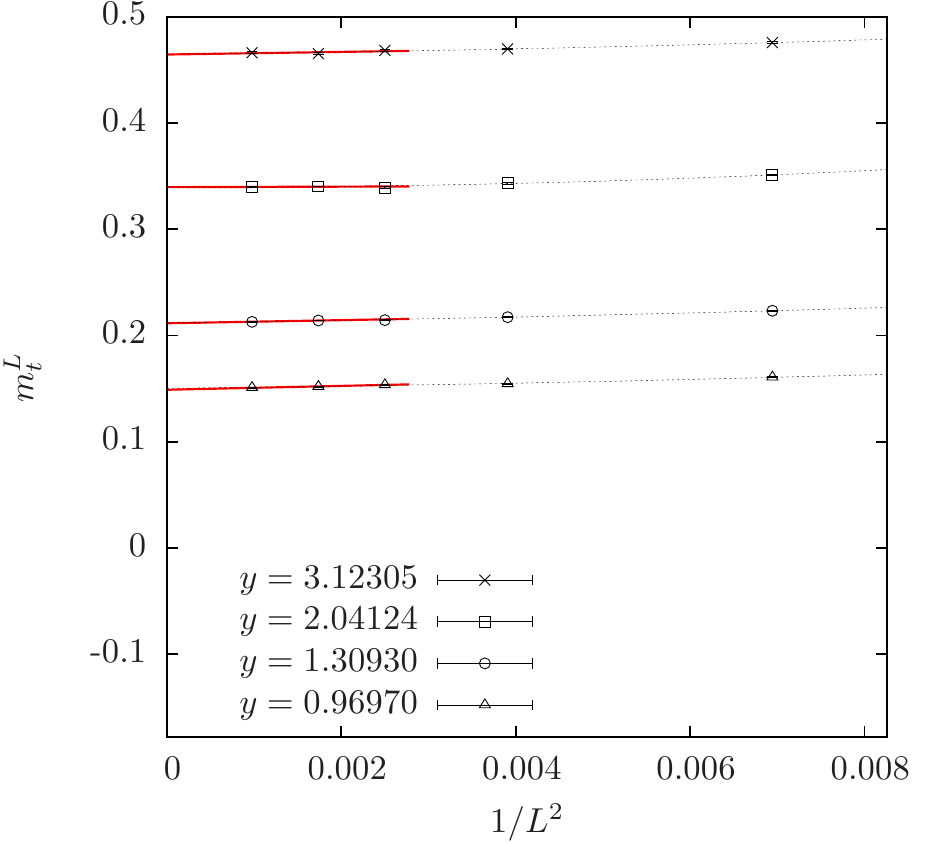} 
						\includegraphics[width=0.31\textwidth]{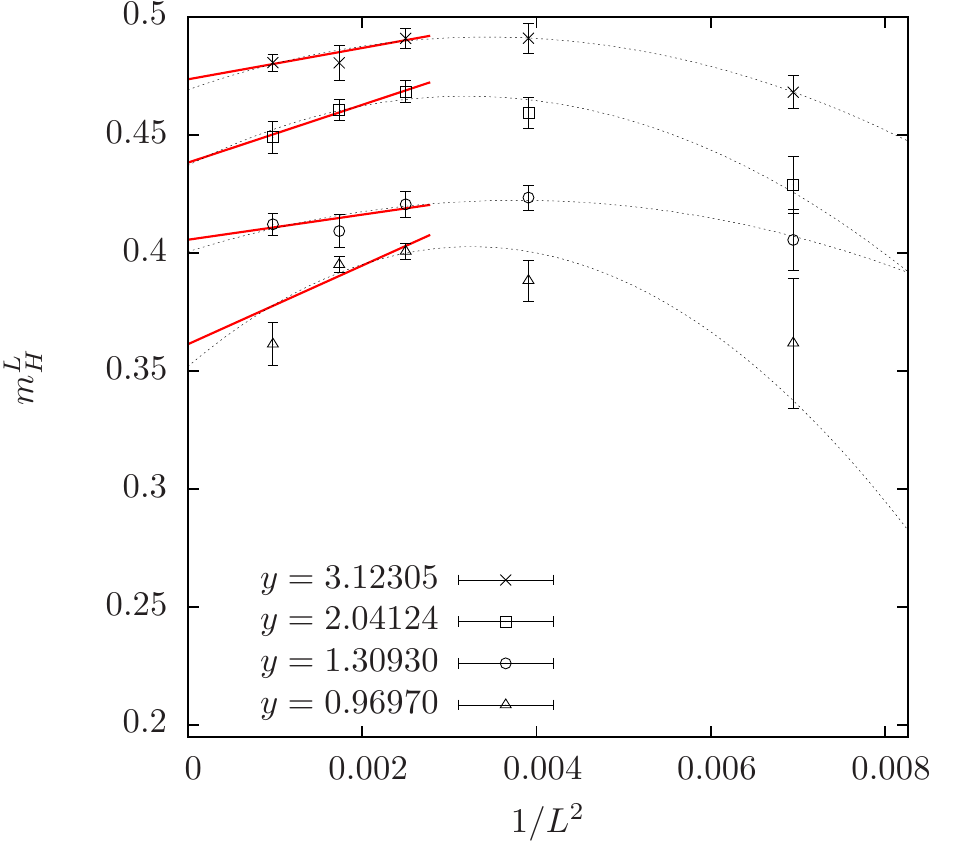}

 						\caption{Finite size effects in (from left to right) the 
						renormalized vacuum expectation value, fermion mass, and Higgs 
						Boson mass at $\Lambda = 1/a = 1.5\mathrm{TeV}$ and 
						$\lambda = \infty$. Data are shown for several values of the bare 
						Yukawa coupling corresponding to fermion masses in the range 
						$m_f = 200...700\mathrm{GeV}$. 
 				While the renormalized vacuum expectation value exhibits asymptotic 
				behavior already at rather small lattice volumes, for a reliable 
				infinite volume extrapolation of the Higgs boson mass, larger volumes 
				are required.}
					\label{fig:FiniteSizeEffects}
				\end{figure}

Particular attention must be paid to finite volume effects in the model. Due to
 the lack of an explicit symmetry breaking parameter in our Lagrangian, we are 
 in 
the so-called $\epsilon$-regime, rather than the $p$-regime where most 
lattice QCD simulations are performed. In the $\epsilon$-regime finite 
volume effects decrease with $L^{-2}$ rather than $\mathrm{e}^{-ML}$ for 
the $p$-regime~\cite{Hasenfratz:1990fu, Hasenfratz:1989pk}. Due to these algebraic finite size effects, an infinite volume extrapolation of the data is necessary. An example of such an extrapolation is shown 
in Fig.~\ref{fig:FiniteSizeEffects}. 

\begin{figure}[ht]
						\centering
						\includegraphics[width=0.4\textwidth]{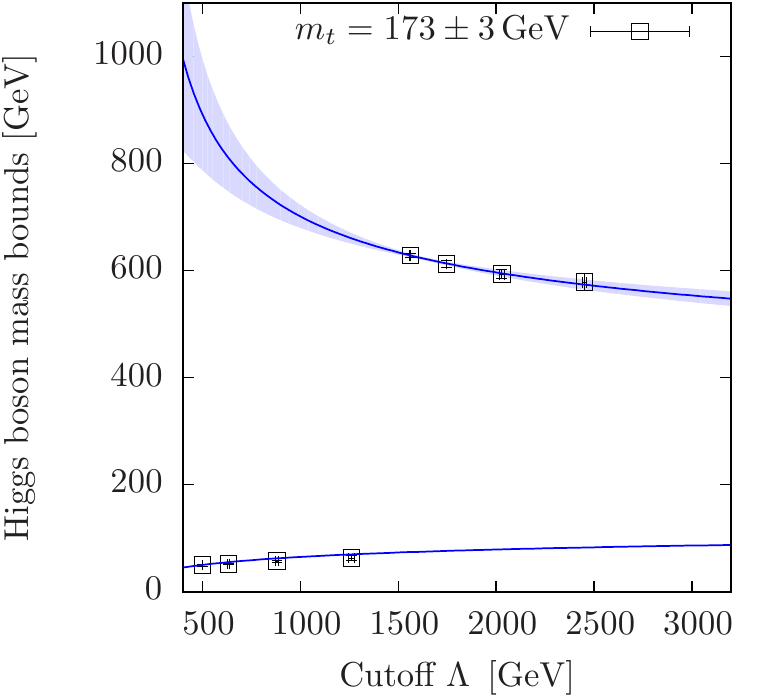}
						\includegraphics[width=0.4\textwidth]{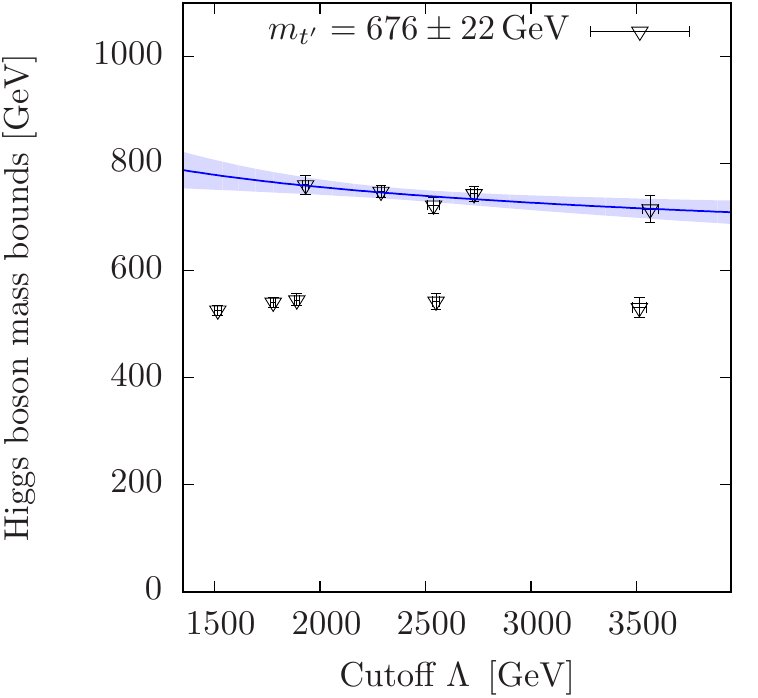} 
 						\caption{The cutoff dependence of the upper and lower Higgs boson 
						mass bounds for the physical top quark mass 
						$m_f = m_t = 175\mathrm{GeV}$ (left) and for 
						$m_f = m_{t'} = 676\mathrm{GeV}$ (right). All data have been 
						extrapolated to infinite volume.} 
						\label{fig:mass_bounds}
\end{figure}

\begin{figure}[ht]
\centering
\includegraphics[width=.4\textwidth]{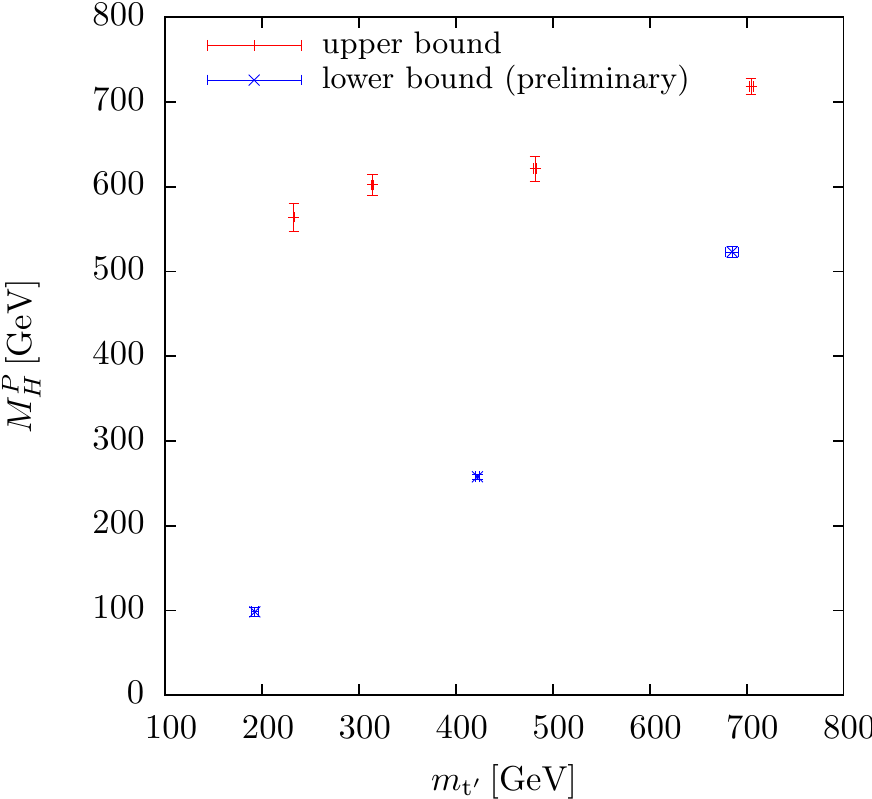}
\caption{Higgs boson mass bounds at $\Lambda = 1/a = 1.5\mathrm{TeV}$ as a 
function of $m_{f} = m_{t'}$. The data for the upper bound has been 
extrapolated to infinite volume, while for the lower bound the results were 
computed on a single $(24^3\times 48)$ volume and are therefore preliminary.}
\label{fig:mass_scan}
\end{figure}

This procedure was first carried out for the situation $m_f = m_t = 175\mathrm{GeV}$~\cite{Gerhold:2010bh} and repeated for $m_f = 676\mathrm{GeV}$~\cite{Gerhold:2010wv} for various values of the lattice cutoff. Results from these two 
calculations are shown in Fig.~\ref{fig:mass_bounds}.  
We now report on additional results from a scan in $m_f$ at fixed cutoff 
$\Lambda = 1/a = 1.5\mathrm{TeV}$. These results are shown in 
Fig.~\ref{fig:mass_scan} and should be regarded as preliminary, as an infinite volume extrapolation for the lower bound has not yet been performed.   

As we discussed above, the method for determining the Higgs boson mass neglects its finite decay width into Goldstone bosons. 
Here we report on a calculation of the resonance parameters of the Higgs boson~\cite{oai:export}, demonstrating that at $m_f = m_t$ the mass obtained by a 
resonance fit to the scattering phase shift is statistically equivalent to the 
pole mass and correlator mass. 

Infinite volume elastic scattering phase shifts may be extracted from 
finite 
volume lattice data by examining the dependence of the energy eigenvalues near 
elastic thresholds~\cite{Luscher:1990ck}. As this technique is valid only for 
elastic scattering, we must add a small explicit symmetry breaking term to the 
Lagrangian, which generates a finite Goldstone boson mass $M_G$. The magnitude 
of the symmetry breaking parameter is chosen such that $M_G \approx M_H/3$. 
Accordingly, 
our results for the scattering phase shift are confined to the region 
$k < 2M_G$.

\begin{figure}[ht]
						\centering
						\includegraphics[width=0.49\textwidth]{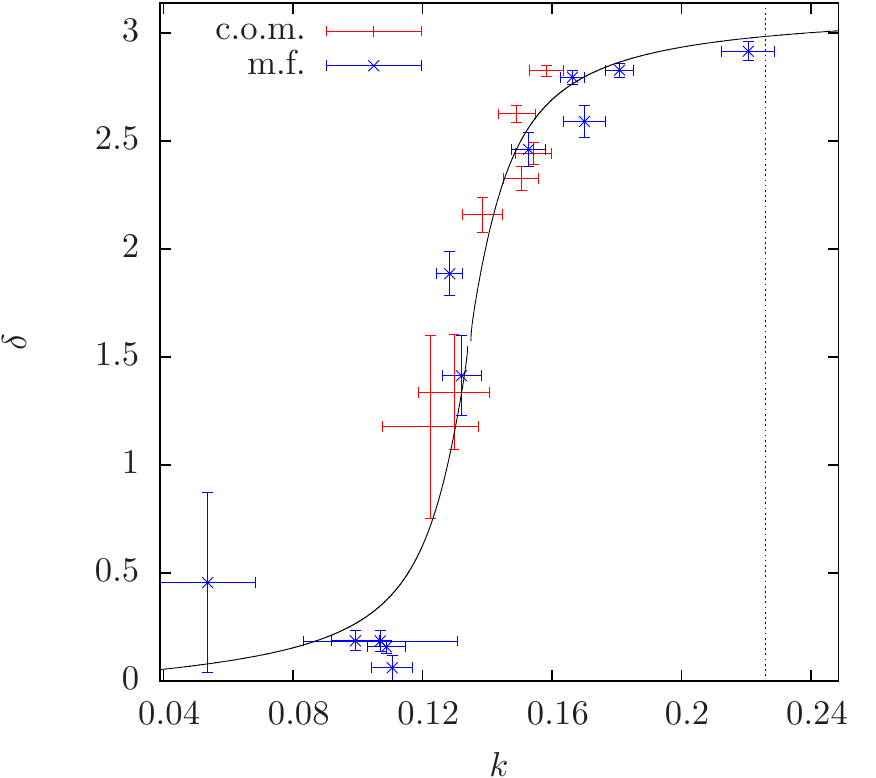} 
						\includegraphics[width=0.49\textwidth]{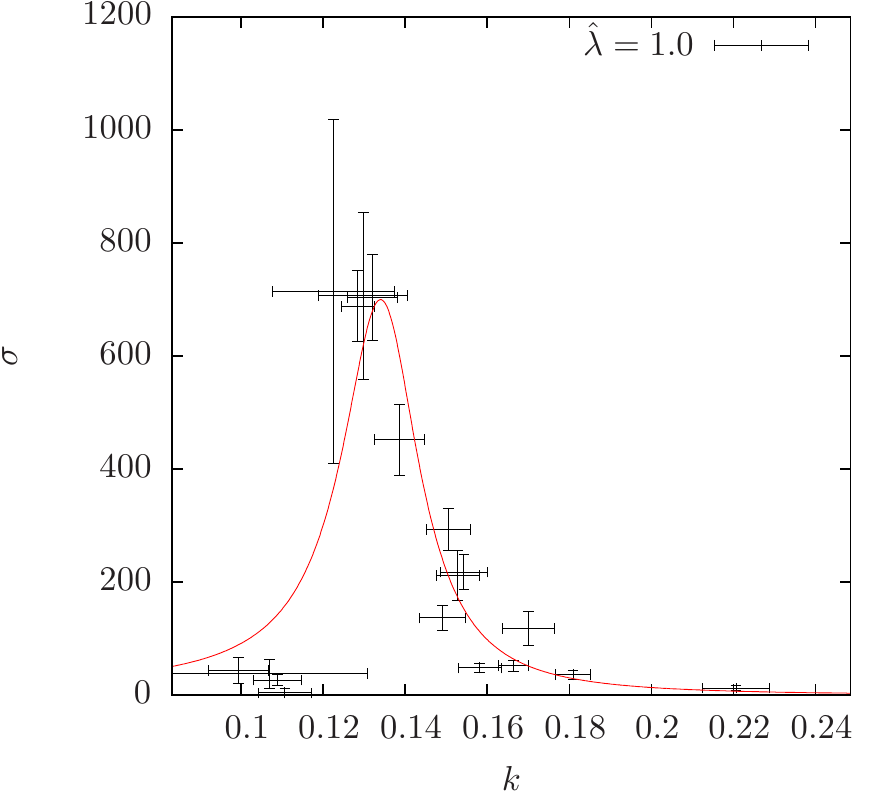} 
						\caption{The scattering phases (left) and the total cross section 
						(right) for $\Lambda \approx 1.5\mathrm{TeV}$ at 
						$m_f = m_t = 175\mathrm{GeV}$ and $\hat \lambda = 1.0$. These 
						results are from various lattice volumes: $L_{s}/a\times40$ with 
						$L_s/a = {12,16,18,20,24,32,40}$. Points obtained using both the center of mass frame (c.o.m) and moving frame (m.f.) are shown. Adding the moving frame is 
						crucial to obtain a reasonable momentum resolution. The vertical 
						line in the left plot denotes the inelastic threshold.}
						\label{fig:phase_example}
\end{figure}

To extract energy eigenvalues of the lattice Hamiltonian, a matrix of 
correlation functions must be constructed and the corresponding generalized eigenvalue problem (GEVP) must be solved~\cite{Blossier:2009kd, Luscher:1990ck}.  
The momentum resolution of the scattering phase shifts is increased by
examining the system at rest and in a moving frame~\cite{Rummukainen:1995vs}. 
Results are shown in Fig.~\ref{fig:phase_example} and Tab.~\ref{tab:resonance}. From this analysis we can see that at least for $m_f = m_t$ and $M_G \approx M_H/3$, the Higgs boson has a relatively narrow width to decay into two Goldstones and determinations of the Higgs boson mass that neglect this decay width are consistent with the full resonance analysis.  

\begin{table}[htb]\centering{
						\begin{tabular}{c c c c c}
						\hline
						$ \hat{\lambda}$  &  Cutoff $\Lambda$=1/a   &  $aM_H^R$           &  $a\Gamma_H^R$   &  $aM_H^P$ \\ \hline
						0.01            & 883(1) GeV           &  0.2811(6)        &  0.007(1)          &  0.278(2)        \\
						1.0             & 1503(5) GeV          &  0.374(4)         &  0.033(4)          &  0.386(28)       \\
						$\infty$        & 1598(2) GeV          &  0.411(3)         &  0.040(4)          &  0.405(4)        \\ \hline
						\end{tabular}}
 						\caption{The resonance mass $M_H^R$ of the Higgs boson together 
						with the resonance width and the mass extracted from 
 						the propagator $M_H^P$ at $m_f = m_t = 175\mathrm{GeV}$. For all three values of the bare quartic coupling, the width is less than 10\% of the resonance mass.}
						\label{tab:resonance}
					\end{table}

Another physical observable that may be studied in this model is 
the finite temperature phase transition from the symmetric phase to the 
broken phase, which may have implications for electroweak 
baryogenesis\cite{Cohen:1993nk}. This phase transition is known to be second order 
in pure $\phi^4$ theory (see e.g. Ref.~\cite{Jansen:1989gd}). However, when SU(2) gauge fields are added, the transition 
becomes first order as in Ref.~\cite{Fodor:1994sj}.  

The temperature in lattice field theory is changed by varying the temporal 
extent of the lattice, with periodic (anti-periodic) temporal boundary 
conditions for the boson (fermion) fields. The phase transition temperature 
is determined by 
first choosing a temporal extent $L_{t}/a$ in lattice units and performing a 
scan in $\kappa$ at fixed $\lambda = \infty$ to determine the peak in the susceptibility. 

\begin{figure}[ht]
\includegraphics[width=0.49\textwidth]{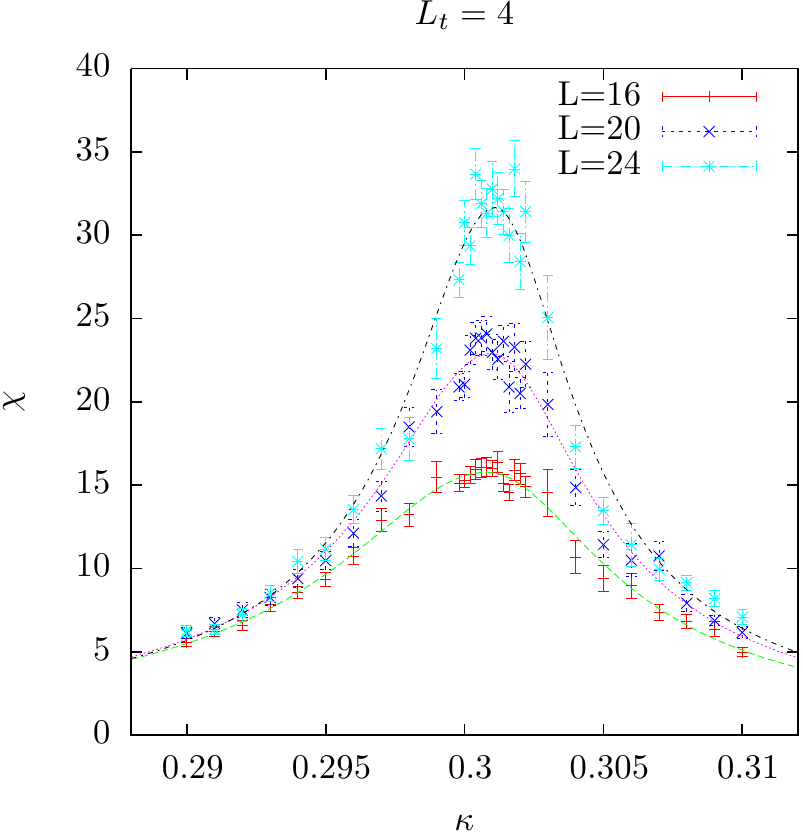}
\includegraphics[width=0.49\textwidth]{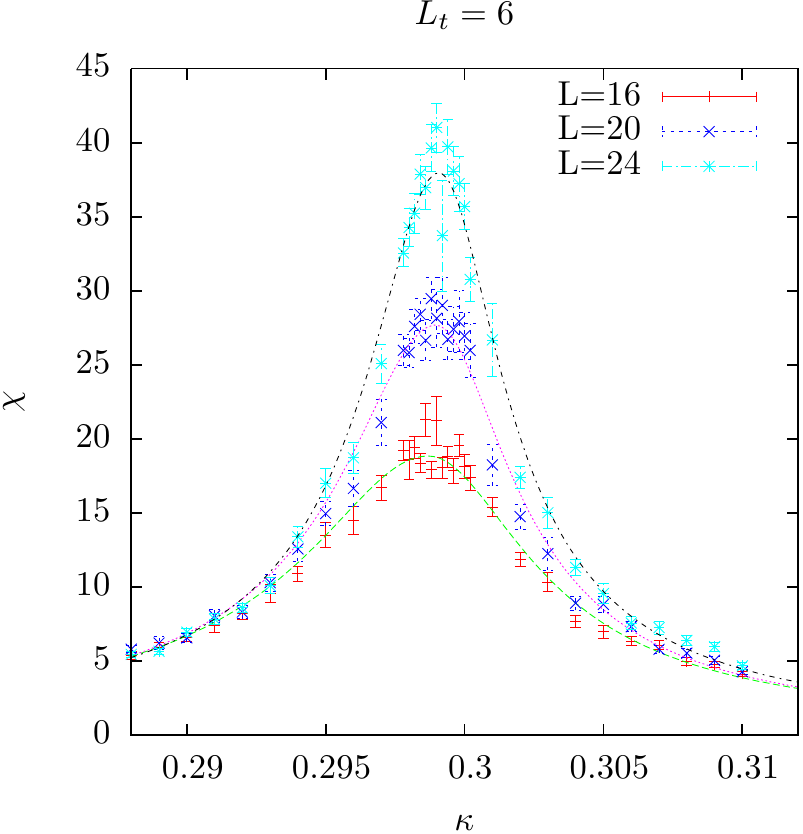}
\caption{The susceptibility as a function of $\kappa$ for 
$m_f = m_t = 175\mathrm{GeV}$. The data shown is for $\lambda = \infty$ and 
should be considered preliminary. The left plot corresponds to $L_t / a  = 4$ 
while the right to $L_t/a = 6$. The dotted lines are from fits to finite size 
and critical scaling ansatz $\chi = A(L_s^{-2/\nu} + B(\kappa - \kappa_c)^{2})^{-\gamma/2}$ where, due to limited 
(preliminary) statistics, we have fixed $\gamma = 1.38$ and $\nu = 0.68$.
The critical temperature determined from the $L_t /a = 4$ lattices is $T_c = 514(15)\mathrm{GeV}$ while from the $L_t / a = 6$ lattices it is $T_c = 491(24)\mathrm{GeV}$. } 
\label{fig:finite_T}
\end{figure}

The value of $\kappa$ at which the susceptibility peaks is then used to 
perform a zero temperature ($L_{t} = \infty$) simulation to set the scale in 
the usual way, from the zero temperature vacuum expectation value. From this 
scale we can determine $T = 1/L_{t}$ in physical units.
Results for the susceptibility and $L_t/a = 4,6$ are shown in 
Fig.~\ref{fig:finite_T}. The preliminary value of the phase 
transition temperature is similar at these two values of the lattice spacing and is $\approx500\mathrm{GeV}$ in comparison to $\approx350\mathrm{GeV}$ in the 
pure 
$\phi^4$ theory. 

In conclusion, studies of several physical quantities in our 
chirally-invariant Higgs-Yukawa model are underway. The dependence of the 
Higgs boson mass bounds on $m_f$ and $a$ has shown that the lower bound 
seems to be sensitive to the mass of the fermion doublet, but rather mildly 
sensitive to the value of the lattice cutoff. We will complete an $m_f$-scan of these bounds in the near future.  
We also plan to repeat the 
determination of the Higgs boson resonance parameters at several values of 
$m_f$.  
Additionally, it may be interesting to look for bound states of Goldstone 
bosons and quarks. 
 The finite temperature phase transition in our model may have implications for 
electroweak baryogenesis. We plan to complete this analysis at several values of $m_f$ and $a$.

\bibliography{lattice}
\bibliographystyle{h-elsevier}

\end{document}